\documentclass[prl,aps,twocolumn]{revtex4-1}
\usepackage{amsmath}
\usepackage{amsfonts}
\usepackage{graphicx}
\usepackage{color}
\usepackage{xcolor}
\usepackage{float}

\begin{document}
\title {Pinning of Andreev bound states  to zero energy in two-dimensional superconductor-semiconductor Rashba heterostructures}
\author{Olesia Dmytruk, Daniel Loss, and Jelena Klinovaja}
\affiliation{Department of Physics, University of Basel, Klingelbergstrasse 82, CH-4056 Basel, Switzerland}
\date{\today}

\begin{abstract}
We consider a two-dimensional electron gas with Rashba spin-orbit interaction (SOI) partially covered by an $s$-wave superconductor, where the uncovered region remains normal but is exposed to an effective Zeeman field applied perpendicular to the plane. We find analytically and numerically Andreev bound states (ABSs) formed in the normal region and show that, due to SOI and by tuning the parameters of the system deeply into the topologically trivial phase, one can reach a regime where  the energy of the lowest ABS becomes pinned close to zero as a function of Zeeman field.  The energy of such an ABS is shown to decay as an inverse power-law in Zeeman field.  We also consider a superconductor-semiconductor heterostructure with a superconducting vortex at the center and in the presence of strong SOI, and find again ABSs that can get pinned close to zero energy in the non-topological phase.
\end{abstract}

\maketitle

{\it Introduction.} Recently, there has been a lot of experimental progress in fabricating heterostructures that are made of a two-dimensional electron gas (2DEG) with strong Rashba SOI proximity coupled to a superconductor~\cite{rev1,rev2,kjaergaard2016quantized,shabani2016two,kjaergaard2017transparent,suominen2017zero,nichele2017scaling,mayer2019superconducting,mayer2019gate,mayer2019phase}. Such heterostructures have attracted a lot of attention both from experimental and theoretical perspectives as they could host topological states, such as Majorana bound states (MBSs)~\cite{alicea2012new,beenakker2013search,sato1,sato2,sau1,sau2,sau3}.  
For example, zero-bias conductance peaks compatible with MBSs were observed in nanowires lithographically defined on  InAs/Al heterostructures~\cite{nichele2017scaling,suominen2017zero}. 
Experiments on one-dimensional setups consisting of a proximitized nanowire with a normal region demonstrated the appearance of zero-bias conductance peaks~\cite{deng2016majorana,vaitiekenas2018effective,deng2018nonlocality,deMoor2018electric}. 
However, in a number of recent theoretical works it was demonstrated that  ABSs can mimic the signatures of MBSs~\cite{abs1,abs2,abs3,abs4,abs5,liu2017andreev,reeg2018zero,liu2019conductance,alspaugh2020volkov,
elsa,abs6,abs7,abs8,abs9,abs10}. Zero-bias peaks coming from the ABSs of  quantum dots in van-der-Waals layers were experimentally observed in Ref.~\cite{dvir2019zeeman}. 
Moreover, MBSs are predicted to appear in genuine 2D systems in  
vortices~\cite{sau2010generic,bjornson2013vortex,bjornson2015probing}, with  zero-bias peaks  observed at the vortex center~\cite{sun2016majorana,exp1,exp2,exp3}.
It is thus of utmost importance to understand the behavior of ABSs in such systems in great detail in order to distinguish them from topological bound states that can potentially emerge in the same setup.

\begin{figure}[t]
\centering
\includegraphics[width=0.35\linewidth]{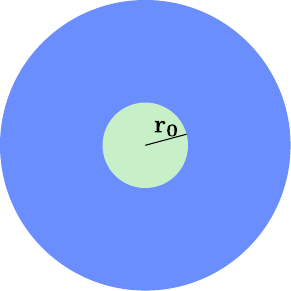}
\caption{Scheme of the system: a 2DEG  with  Rashba SOI.  The outer part of the disk (blue) is covered by a bulk $s$-wave superconductor resulting in the proximity-induced pairing potential $\Delta$. The inner part of the disk (green) is left uncovered by the superconductor, forming a normal part of  radius $r_0$, hosting the ABS. 
The Zeeman field, produced by e.g. an insulating ferromagnet~\cite{sau1} (not shown), is applied to the normal region and points perpendicular to the plane.}
\label{fig:setup}
\end{figure}

In this work, we focus on ABSs occurring in a confined 2DEG region with strong Rashba SOI partially proximitized with an $s$-wave superconductor, see Fig.~\ref{fig:setup}. A Zeeman field is applied to the normal region perpendicular to the plane. Such effective Zeeman fields  can be generated by placing an insulating ferromagnet on top of the normal region with proximity-induced exchange coupling~\cite{sau1,sau3}.
The ABSs are induced in the normal part of the system, which is left uncovered by the superconductor. We perform both  analytical and numerical calculations of the ABS spectrum for two different 2D geometries: a disk and a rectangular lattice. We find that ABSs can be pinned close to zero energy as a function of Zeeman field in the topologically trivial phase by tuning the SOI length with respect to the typical size of the normal part. Moreover, we study a 2D heterostructure with a superconducting vortex, and demonstrate that there are ABSs at the vortex core due to SOI pinned close to zero-energy, again, in the topologically trivial phase.

{\it Analytical  model.} We consider a 2DEG with standard 2D Rashba SOI.  The finite-size system, for simplicity, is assumed to have disk shape, the outer part of which is covered by an $s$-wave superconductor. In contrast to that, the inner part of the disk is left uncovered, thus, forming the normal disk region of  radius $r_0$.  In addition, the inner part is subjected to an effective Zeeman  field. For  sake of simplicity, here, in the analytical part of the paper, we neglect  the Zeeman energy and SOI in the outer part. Later, we show numerically that this assumption is not crucial for the observation of the proposed effects.
%and the proposed effects can still be observed.
%even if it is not fulfilled. 
The chemical potential $\mu_s$ of the outer part is also assumed much larger than that of the normal part, where we assume the latter to be tuned to the SOI energy. Such a model is consistent with the metallization effects caused by the bulk superconductor, where strong coupling between super- and semiconductor can substantially modify the semiconducting properties of the sample~\cite{reeg2017finite,reeg2018mettalization,reeg2018proximity,mikkelsen2018hybridization,woods2018effective,antipov2018effects}. 
We note that the topological phase with MBSs cannot be reached in our system due to the large shift of the chemical potential $\mu_s$. 
Instead, however, we will find  ABSs that become pinned close to zero energy in the middle of the superconducting gap for sufficiently large Zeeman fields.

Making use of the rotation invariance of the setup, we work in polar coordinates, $ x = r\cos\varphi$ and $y = r\sin\varphi$ with $\varphi \in [0,2\pi)$ and $r>0$. The Hamiltonian of the normal part $\mathcal{H}_n(r,\varphi)$ defined for $0 < r < r_0$ and of the superconducting part $\mathcal{H}_s(r,\varphi) $ defined for $r>r_0$ are given by 
\begin{align}
&\mathcal{H}_n(r,\varphi) = -\dfrac{\hbar^2}{2m^*} \left(\partial^2_{r}  + \dfrac{1}{r}  \partial_{r} + \dfrac{1}{r^2} \partial^2_{\varphi}\right) \eta_z\notag\\
&\hspace{30pt}- i\alpha e^{- i \varphi \eta_z \sigma_z/2} \eta_z \sigma_y e^{i \varphi \eta_z \sigma_z/2} \partial_{r}\notag\\
&\hspace{30pt}+ i\dfrac{\alpha}{r} e^{- i \varphi \eta_z \sigma_z/2} \sigma_x e^{i \varphi  \eta_z \sigma_z/2} \partial_{\varphi}+\Delta_Z \eta_z\sigma_z,\\
&\mathcal{H}_s(r,\varphi) = -\dfrac{\hbar^2}{2m^*_s} \left(\partial^2_{r}  + \dfrac{1}{r} \partial_{r} + \dfrac{1}{r^2} \partial^2_{\varphi}\right) \eta_z \notag\\
&\hspace{30pt}-\mu_s \eta_z + \Delta \eta_y\sigma_y.
\end{align}
respectively. In what follows, we use the basis $ \left(\psi_{\uparrow},\psi_{\downarrow},\psi^\dag_{\uparrow},\psi^\dag_{\downarrow}\right)$, where $\psi_{\sigma}$ is a standard annihilation operator acting on the electron with spin $\sigma$.
Here, $m^*$ ($m^*_s$) is the effective mass in the normal (superconducting) part, $\alpha$ is the Rashba SOI strength, 
$\Delta_{Z}$ is the Zeeman energy, $\mu_s$  the chemical potential in the superconducting part, and $\Delta$  the proximity-induced superconducting pairing potential. Here, the Pauli matrices $\sigma_{x,y,z}$  ($\eta_{x,y,z}$) act in spin (particle-hole) space.

Next, we note that the total angular momentum defined by the operator $\hat J_z = - i\hbar \partial_\varphi + \eta_z\sigma_z/2$ is conserved~\cite{tsitsishvili2004rashba,yang2016majorana,garnier2019topological,volpez2019second}. Thus, the wavefunctions can be written in the  form $\Psi_m(r,\varphi) = e^{- i \eta_z\sigma_z \varphi/2} e^{im \varphi} R_m(r)$,
where $m$ are the eigenvalues of $\hat J_z$ and $R_m (r) = [u_{m,\uparrow}(r), u_{m, \downarrow}(r), v_{m,\uparrow}(r), v_{m,\downarrow}(r)]^T$ is the radial part of the wavefunction.
To ensure single-valuedness with periodic boundary conditions, $\Psi_m(r,\varphi) = \Psi_m(r,\varphi+2\pi)$, only half-integer values of $m$ are allowed. The particle-hole symmetry is restored if both blocks with $m$ and $-m$ are taken into account.
After the transformation the resulting Hamiltonians for a given block $m$ become
\begin{align}
\mathcal{H}^{m}_n(r) &= -\dfrac{\hbar^2}{2m^*} \left[\partial^2_{r}  + \dfrac{1}{r}  \partial_{r} -\dfrac{\left(m-\eta_z\sigma_z/2\right)^2}{r^2} \right] \eta_z\label{eq:BdGnormal} \notag\\
&- i\alpha \eta_z  \sigma_y  \left(\partial_{r} + \dfrac{1}{2r}\right)- \dfrac{\alpha m}{r} \sigma_x    +\Delta_Z \eta_z\sigma_z,\\
\mathcal{H}^{m}_s(r) &= -\dfrac{\hbar^2}{2m^*_s} \left[\partial^2_{r}  + \dfrac{1}{r}  \partial_{r}  - \dfrac{\left(m-\eta_z\sigma_z/2\right)^2}{r^2} \right] \eta_z  \notag\\
&-\mu_s\eta_z+ \Delta \eta_y\sigma_y.
\label{eq:BdGsuper}
\end{align}
The eigenvalue equations are given by $\mathcal{H}^{m}_n(r) R_m^{(n)}(r)$ $=$ $E R^{(n)}_m(r)$ and $\mathcal{H}^{m}_s(r) R_m^{(s)}(r)$ $=$ $E R^{(s)}_m(r)$.  %\jk{a bit strange here, do we mean: $\mathcal{H}^{m}_n(r) R_m^{(n)}(r) = E R^{(n)}_m(r)$} 
In addition, the corresponding wavefunction should satisfy matching boundary conditions at $r=r_0$.

{\it Andreev bound state energy.} In a  next step, we find the energy $E$ ($E< \Delta$) of the energetically lowest-lying ABS analytically. 
The wavefunction in the normal part is determined from Eq.~\eqref{eq:BdGnormal} by using a wavefunction ansatz in  form of Bessel functions~\cite{tsitsishvili2004rashba}. We restrict ourselves to the solutions that can be normalized at $r = 0$  and focus on the strong SOI regime, $E_{so} \gg \Delta_Z$, where 
$E_{so}=m\alpha^2/(2\hbar^2)$  is the SOI energy \cite{comp}. The wavefunction in the superconducting part is determined by  Eq.~\eqref{eq:BdGsuper} in the regime of large chemical potential, $\mu_s \gg \Delta$, where only decaying solutions are taken into account. 
The corresponding wavefunctions read 
\begin{widetext}
\begin{align}
&R_m^{(n)}= 
c_1 \begin{pmatrix}
 i\gamma J_{m-\frac{1}{2}}(k^{(n)} r)  \\
J_{m+\frac{1}{2}}(k^{(n)} r)   \\
0\\
0
\end{pmatrix}+
c_2 \begin{pmatrix}
J_{m-\frac{1}{2}}(k_{+}^{(n)} r)  \\
J_{m+\frac{1}{2}}(k_{+}^{(n)} r)  \\
0\\
0
\end{pmatrix}
+
c_3 \begin{pmatrix}
0\\
0\\
\left(i/\gamma\right) J_{m+\frac{1}{2}}(k^{(n)} r)   \\
-J_{m-\frac{1}{2}}(k^{(n)} r) 
\end{pmatrix}+
c_4 \begin{pmatrix}
0\\
0\\
 -J_{m+\frac{1}{2}}(k_{-}^{(n)} r)  \\
J_{m-\frac{1}{2}}(k_{-}^{(n)} r) 
\end{pmatrix},\\
&R_m^{(s)} = 
c_5 \begin{pmatrix}
e^{i\phi}  K_{m-\frac{1}{2}}(k_{-}^{(s)} r)  \\
 0\\
 0\\
-K_{m-\frac{1}{2}}(k_{-}^{(s)} r)  
\end{pmatrix}+
c_6 \begin{pmatrix}
e^{-i\phi}  K_{m-\frac{1}{2}}(k_{+}^{(s)} r)   \\
0\\
0\\
-K_{m-\frac{1}{2}}(k_{+}^{(s)} r)  
\end{pmatrix}
+
c_7 \begin{pmatrix}
0\\
e^{i\phi}  K_{m+\frac{1}{2}}(k_{-}^{(s)} r)  \\
K_{m+\frac{1}{2}}(k_{-}^{(s)} r)  \\
0
\end{pmatrix}+
c_8 \begin{pmatrix}
0\\
e^{-i\phi}  K_{m+\frac{1}{2}}(k_{+}^{(s)} r)  \\
K_{m+\frac{1}{2}}(k_{+}^{(s)} r)   \\
0
\end{pmatrix},
\label{eq:normalnew}
\end{align}
\end{widetext}
where $\gamma=\sqrt{\left(\Delta_Z+E\right)/\left(\Delta_Z-E\right)}$, $k^{(n)} \approx i \sqrt{\Delta_Z^2 - E^2}/\alpha$, $k_{\pm}^{(n)} \approx \left(2 k_{so} \pm  E/\alpha\right)$ with $k_{so} = m^*\alpha/\hbar^2$ being the SOI momentum, $\cos\phi = E/\Delta$, $k_{\pm}^{(s)} \approx \pm i k_F+\sqrt{\Delta^2 - E^2}/\left(\hbar v_F\right) $ with $k_F = \sqrt{2m_s^*\mu_s}/\hbar$ being the Fermi momentum and $v_F = \sqrt{2\mu_s/m_s^*}$ the Fermi velocity in the superconductor. Here, $J_\lambda(x)$ is the Bessel function of first kind of order $\lambda$~\cite{BesselJ} and  $K_\lambda(x)$ is the modified Bessel function of  second kind of order $\lambda$~\cite{BesselK}.

\begin{figure}[t!]
\centering
\includegraphics[width=\linewidth]{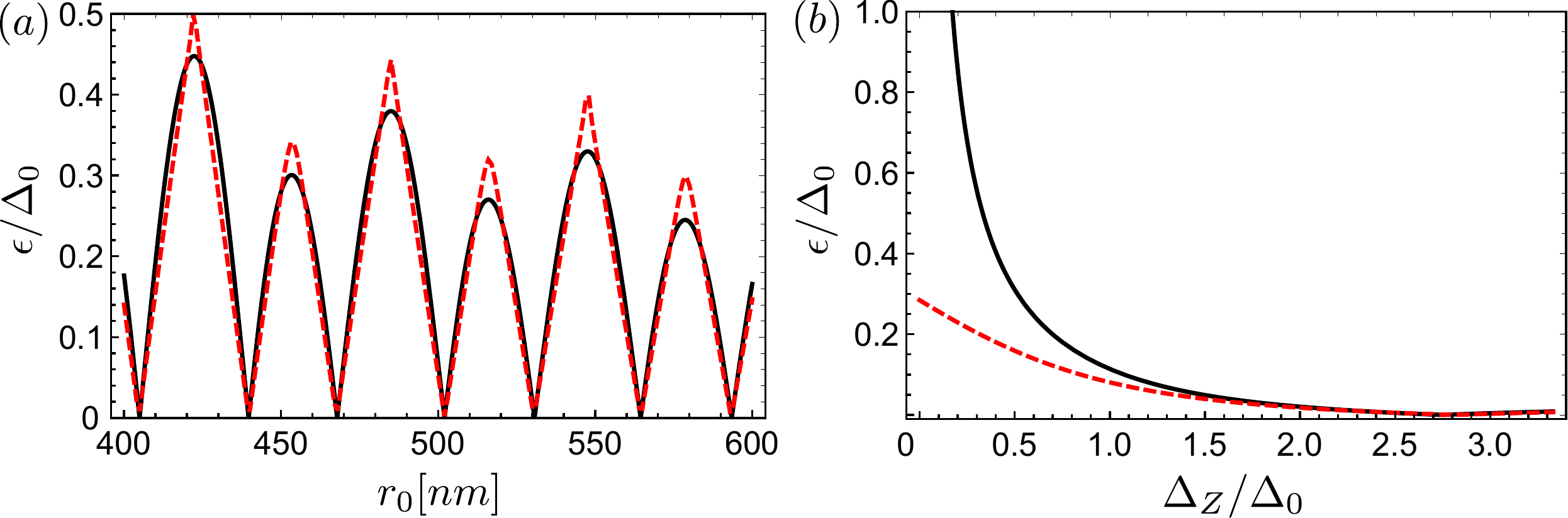}
\caption{(a) Energy $\epsilon$ of the lowest ABS  found analytically in  Eq.~\eqref{eq:energyapp} (black solid line) and found numerically by solving $\det M_m(\epsilon) = 0$ (red dashed line) as a function of the normal part radius $r_0$ for $m = 1/2$  
and  $\Delta_Z=2.7\Delta_0$.
The ABS energy oscillates and vanishes at specific values of the radius $r_0$. (b) As we tune the radius to one of these $\epsilon=0$ values (here, $r_0 = 502$~nm), the lowest ABS is pinned near zero energy over a wide range of Zeeman energy $\Delta_Z$. 
There is  good agreement between numerical and analytical solutions for $\epsilon \ll \Delta_Z$ and $\alpha/r_0 \ll \Delta_Z \ll E_{so}$.
 Other parameters are chosen as $\Delta_0 = 0.2$~meV, $E_{so} = 0.8$~meV, $\hbar v_F = 10 \alpha$, $m^* = 0.03 m_e$, $\mu_n = 0$.}
\label{fig:spectrumformula}
\end{figure}

To find the energy of the ABS, we impose two boundary conditions. First, we impose that the wavefunction should be continuous at $r = r_0$: $R_m^{(n)}(r_0)=R_m^{(s)}(r_0)$.
 Second, the quasiparticle current at $r= r_0$ should be conserved, $\hat{v}_nR_m^{(n)}(r_0)= \hat{v}_sR_m^{(s)}(r_0)$ \cite{boun1,boun2}. 
Introducing the radial momentum operator  $\hat p=-i \hbar\left[\partial_r + 1/\left(2r\right)\right]$~\cite{fujikawa2008non,yang2016majorana} and defining $\hat{v}_{n,s} = \partial \mathcal{H}_{n,s}/\partial \hat{p}$, we find that $\hat{v}_n$ $=$  $- i \left(\hbar/m^*\right) \left[\partial_r + 1/\left(2r\right)\right]\eta_z + \left(\alpha/\hbar\right) \eta_z\sigma_y$ and $\hat{v}_s$ $=$ $- i\left(\hbar/m_s^*\right) \left[\partial_r + 1/\left(2r\right)\right]\eta_z$. These two boundary conditions can be arranged as a single algebraic equation of the form $M_m\vec{c} = 0$, where $\vec{c}$  is a vector of unknown coefficients $c_{1},\ldots, c_{8}$ and $M_m$ is an $8\times 8$ matrix. This matrix equation has a nontrivial solution if $\det M_m = 0$, from which one determines the energy $\epsilon$ of the lowest ABS.
  Next, expanding the determinant up to  first order in $E$ in the limit $\Delta_Z \gg \alpha/r_0$ and keeping only the terms proportional to the highest order in $\mu_s$, we obtain the following expression for the energy of the lowest ABS  for $m = 1/2$:
\begin{align}
&\epsilon \approx \dfrac{\alpha}{2r_0 \left(\left[J_0(2k_{so}r_0)\right]^2+\left[J_1(2k_{so}r_0)\right]^2\right)} \Big[\left[J_0(2 k_{so}r_0 )\right]^2\notag\\
&\times\dfrac{  I_1\left(\Delta_Z r_0/\alpha\right)}{  I_0\left(\Delta_Z r_0/\alpha\right) }
- \left[J_1(2 k_{so}r_0 )\right]^2\dfrac{  I_0\left(\Delta_Z r_0/\alpha\right)}{ I_1\left(\Delta_Z r_0/\alpha \right)}\Big],
\label{eq:energyapp}
\end{align}
where  $I_\lambda(x)$ is the modified Bessel function of first kind of  order $\lambda$~\cite{BesselI}. Note that $\epsilon$
%the energy $\epsilon$ of the ABS 
given by Eq.~\eqref{eq:energyapp} depends only on the parameters of the normal part. Moreover, the Bessel functions $J_{0,1}(2k_{so}r_0)$ cross zero at specific values of $k_{so}r_0$. Tuning this product to one of these values at large values of the Zeeman field will allow us to find the ABS lying close to zero energy.
The ratio $ I_0\left(\Delta_Z R/\alpha\right)/ I_1\left(\Delta_Z R/\alpha \right)$ decays as a power-law with increasing Zeeman field, thus ensuring the possibility of having a near zero-energy ABS in the system.

We compare our analytical expression given by Eq.~\eqref{eq:energyapp} with the ABS spectrum found numerically by solving the equation $\det M_m(\epsilon) = 0$. 
%In what follows, we also take into account that, if the Zeeman field $B$ is applied globally, the pairing potential could be suppressed as a function of the external Zeeman field $\Delta = \Delta_0\sqrt{1-\left(B/B_c\right)^2}$, where $B_c$ is the critical field at which superconductivity is destroyed. As already follows from Eq.~(\ref{eq:energyapp}), this assumption in not crucial as long as one is below $B_c$. 
First, we fix the value of the Zeeman energy and calculate the energy of the lowest ABS as a function of the normal part radius $r_0$ [see Fig.~\ref{fig:spectrumformula}(a)]. The ABS energy oscillates as a function of $r_0$ passing through zero only for specific values of $r_0$. Next, we fix  $r_0$ to one of these values and calculate the lowest ABS energy $\epsilon$ as a function of the Zeeman energy $\Delta_Z$. 
We find that $\epsilon$
stays close to zero over a large range of $\Delta_Z$ [see Fig.~\ref{fig:spectrumformula}(b)]. 
In general, the ABS energy $\epsilon$ is pinned close to zero for all values of $r_0$ at which $\epsilon$ goes to zero in Fig.~\ref{fig:spectrumformula}(a), or, more generally, at which the energy separation $\alpha/r_0 \ll \Delta_Z \ll E_{so}$ [this comes from the conditions imposed during the derivation of the wavefunctions and Eq.~\eqref{eq:energyapp}] is respected. Moreover, there is  good agreement between the numerical solution and the analytical expression for $\epsilon$ given by Eq.~\eqref{eq:energyapp} for $\epsilon \ll \Delta_Z$.

\begin{figure}[t]
\centering
\includegraphics[width=\linewidth]{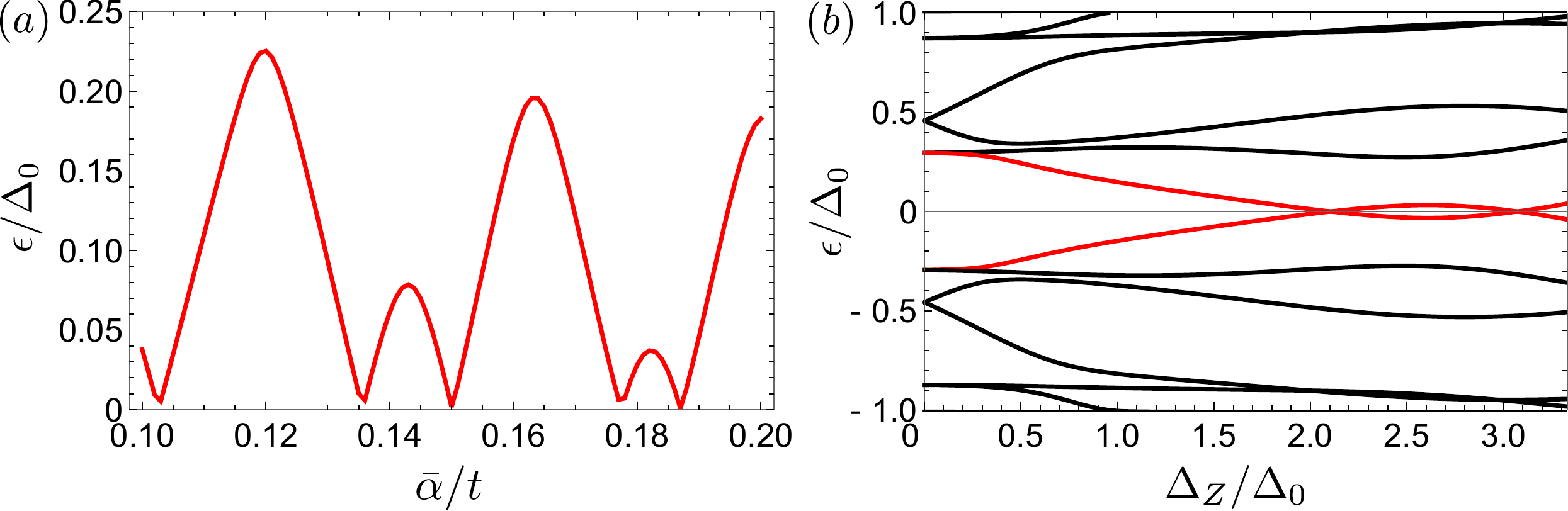}
\caption{(a) Energy $\epsilon$ of the lowest ABS, found numerically by solving the eigenvalue equations for the radial components of the wavefunctions using finite difference methods, oscillating as function of SOI parameter $\bar\alpha/t$ [$\bar\alpha = \alpha/(2a)$]. We fixed $m=1/2$, $\Delta_Z = 2.7\Delta_0$, and assumed a uniform SOI in the entire system.
%normal and superconducting parts. 
The lowest ABS energy vanishes at specific values of $\bar\alpha/t$. 
b) When the SOI is tuned to one of these values, $\bar\alpha/t = 0.184$, the lowest ABS (red curve) is pinned close to zero energy over a large range of Zeeman energy $\Delta_Z/\Delta_0$.
Other parameters are fixed as $\Delta_0 = 0.2$~meV, $\mu_n = 0$, $\mu_s=5 \Delta_0$, $m^* = 0.03 m_e$, $a = 5$~nm ($t= 50$ meV), $r_0/a = 99$, and the length of the outer part $L_{s}/a = 900$.}
\label{fig:spectrumnumerics}
\end{figure}

\begin{figure}[t]
\includegraphics[width=\linewidth]{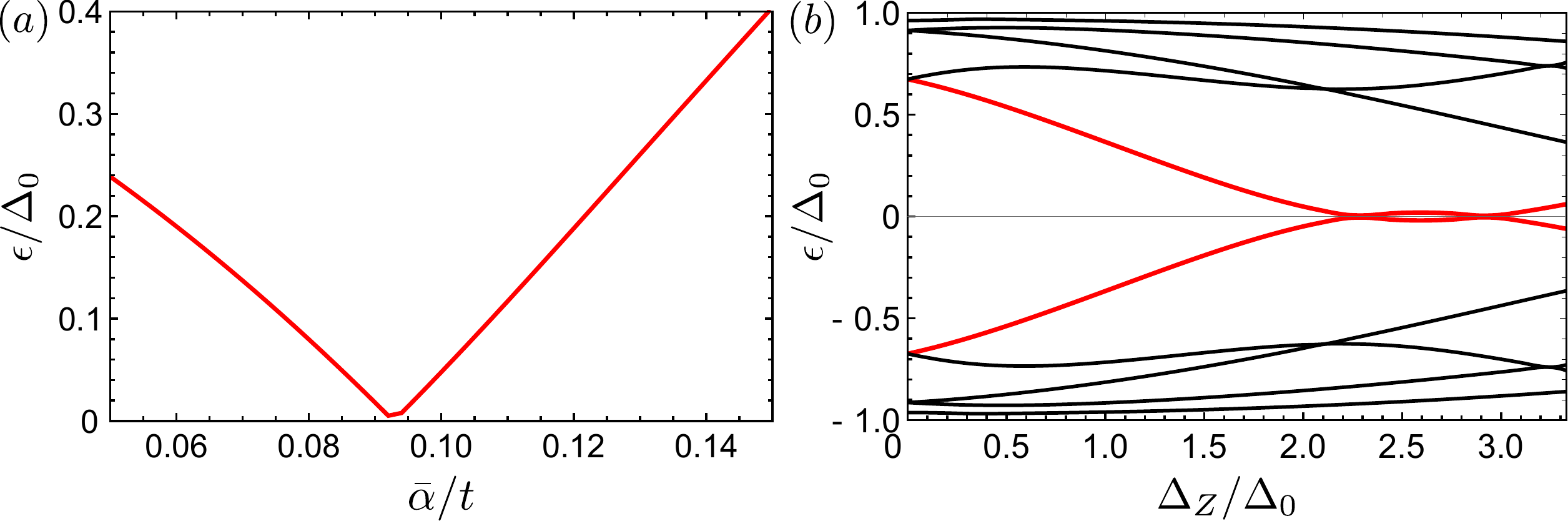}
\centering
\caption{(a) Lowest ABS energy $\epsilon$ of a 2D system on a square lattice approaches zero for a particular value of the SOI $\bar\alpha/t$: here, $\Delta_Z = 3\Delta_0$.
(b) Fixing the SOI to this value, $\bar{\alpha}/t  = 0.094$, we find  the full energy spectrum of the system as a function of Zeeman energy $\Delta_Z/\Delta_0$. The energy of the lowest ABS (red curve) is pinned close to zero energy over a large range of Zeeman fields.
Both plots were obtained by numerically diagonalizing the Hamiltonian on a 2D square lattice with uniform SOI
for $\Delta_0 = 0.2$~meV,  $\mu_n = 0$, $\mu_s = 3.5\Delta_0$, $a = 10$~nm, $N_{x /y} = 200$, 
$N_{x/y,n} = 20$, and $m^*=0.03 m_e$ ($t =13$~meV). 
}
\label{fig:spectrumtight}
\end{figure}

\begin{figure*}[!t]
\centering
\includegraphics[width=\textwidth]{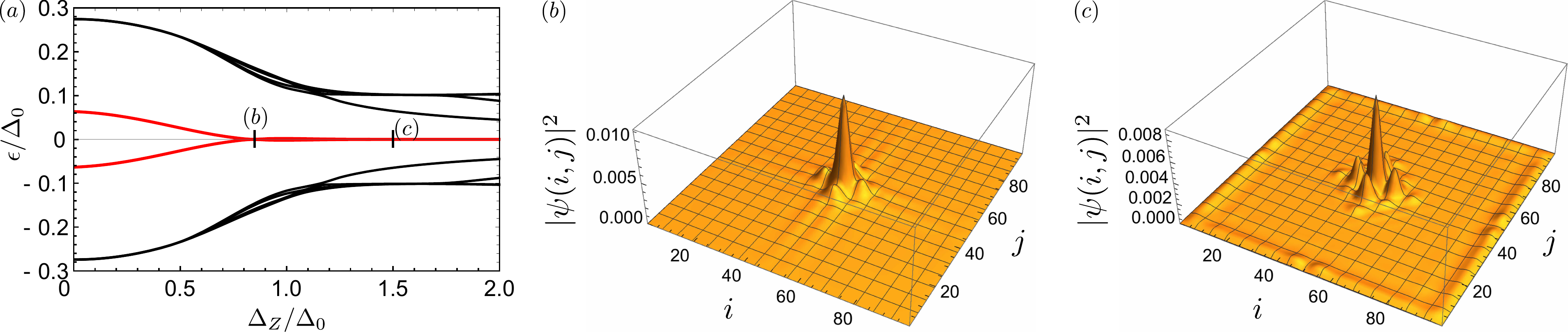}
\caption{(a) Energy spectrum of a 2D system in the presence of the vortex as function of Zeeman energy $\Delta_Z/\Delta_0$. (b) Probability density of the ABS on the site $(i,j)$ in topologically trivial phase [$\Delta_Z = 0.85 \Delta_0$]. The wavefunction of the lowest energy state is localized at the center of the system. (c) Probability density of the MBS on the site $(i,j)$ in topological phase [$\Delta_Z = 1.5 \Delta_0$].  The wavefunction has components both at the center and at the edge of the system. Other parameters are fixed as $N_x = N_y = 99$, $i_0 = j_0 = 50$, $R/a = 2$, $\Delta_0/t = 0.1$, $\overline\alpha/t = 0.28$, $\mu = 0$.
}
\label{fig:vortex}
\end{figure*}

{\it Numerical solution.} Alternatively, to determine the ABS energy, we can solve the eigenvalue equations for the radial components of the wavefunctions numerically. We introduce $\Phi_m(r) = \sqrt{r} R_m(r)$ in Eqs.~\eqref{eq:BdGnormal} and \eqref{eq:BdGsuper} and use the finite difference method with the Neumann  boundary condition at $r = 0$, $\partial_{r}\Phi_m^{(n)}(0) = 0$. 
Again, to find the parameters of the system for which the ABS can be pinned close to zero energy, we first fix the values of the Zeeman field and calculate, this time, the energy of the lowest ABS as a function of  $\alpha$. This scenario is close to realistic experimental setups where by tuning electric gates not only  the chemical potential in the normal part $\mu_n$ is shifted but also the SOI value $\alpha$. Generally, as suggested from Eq.~\eqref{eq:energyapp}, one can tune to the resonance in both ways by changing the effective size $r_0$ or by tuning $\alpha$ or $\mu_n$. We find that the energy of the ABS oscillates as a function of SOI, crossing zero only at specific values of $\alpha$ [see Fig.~\ref{fig:spectrumnumerics}(a)]. Next, we fix the value of $\alpha$ to one of these specific values and calculate the full spectrum as a function of  Zeeman field. We find that for this specific values of $\alpha$ the lowest ABS is pinned close to zero energy over a large range of Zeeman field, see Fig.~\ref{fig:spectrumnumerics}(b). The tiny splitting around zero can be easily masked by line broadening due to temperature effects or coupling the system to transport probes~\cite{broad1,broad2,broad3,broad4,broad5,broad6}.
%by temperature broadening effects and/or by coupling the system to transport probes.
Note that here we consider uniform SOI but we verified that the results remain qualitatively the same when the SOI is set to zero in the superconducting part of the disk as well as we verified that the results do not depend on the assumption of parameters changing as a step function.
Disorder  within the superconducting part of the system does not affect the ABS spectrum even if on-site fluctuations in the chemical potential are larger than $\mu_s$. The lowest ABS remains pinned close to zero energy.

{\it Rectangular lattice.} In the previous sections, we exploited the rotational invariance in a disk geometry. To demonstrate that this is not crucial, we break this symmetry and study a 2D heterostructure on a rectangular lattice.  We also assume that the inner normal part  has a rectangular shape ($N_{x/y,n}$) as well as the system itself ($N_{x,y} $). 
Following the same procedure as before, we calculate the energy of the lowest ABS as a function of the SOI. Again, we find that the lowest ABS energy goes close to zero only for particular values of SOI, see Fig.~\ref{fig:spectrumtight}(a).
Afterwards, we fix $\alpha$ to this value and calculate the full energy spectrum of a 2D system as a function of $\Delta_Z$. The energy of the lowest ABS is pinned near zero for a wide range of Zeeman fields, see Fig.~\ref{fig:spectrumtight}(b).  Also in this geometry, we considered different perturbations, for example,   different locations of the normal part within a 2D lattice, and we find that the ABS stays pinned close to zero energy independent of the location of the normal part for the superconducting coherence length being smaller than the size of the superconductor part.

{\it Bound states in a vortex.} Next, we study the formation of ABSs in superconductor vortices in 2D hybrid structures. 
We consider a 2D system on a square lattice with the superconducting pairing amplitude assumed as $\Delta = 0$ for $r<r_0$ and $\Delta = \Delta_0 e^{i\varphi}$ for $r \geq r_0$, where $r_0$ is  the vortex radius  defined from the center of the system $(i_0, j_0)$, with $i_0 = (N_x+1)/2$ and $j_0 = (N_y+1)/2$, and $\varphi$ is the polar angle.
Formation of MBSs has been  predicted for such systems if $\Delta_Z^2 > \mu^2 + \Delta_0^2$~\cite{sau2010generic}, with one MBS localized at the center and its partner at the system edge.  Again, calculating the energy spectrum of a 2D system with a vortex numerically, we find that there is an ABS pinned close to zero energy due to SOI appearing in the topologically trivial phase defined as  $\Delta_Z^2 < \mu^2 + \Delta_0^2$ [see Fig.~\ref{fig:vortex}(a)]. The probability density of the lowest energy state has the most weight concentrated in the middle of the vortex for $\Delta_Z = 0.85 \Delta_0$ [see Fig.~\ref{fig:vortex}(b)], while the probability density is concentrated both at the center and at the edge of the system in the topological regime for $\Delta_Z = 1.5\Delta_0$ [see Fig.~\ref{fig:vortex}(c)]. Thus, due to SOI there is an ABS pinned close to zero energy forming at the center of the vortex in the topologically trivial phase.

{\it Conclusions.---}In this paper, we studied a 2D system with Rashba SOI partially coupled to an $s$-wave superconductor in the presence of a Zeeman field.
We calculated the energy spectrum of such a system in a disk geometry both numerically and analytically, and we found that the energy of the lowest ABS can be pinned close to zero over a large range of Zeeman fields due to SOI. Due to a large shift of the chemical potential in the superconducting part, the pinning of the lowest ABS depends only on the parameters of the normal part and can be observed by fine-tuning the SOI with respect to the normal part radius if the energy separation $\alpha/r_0 \ll \Delta_Z \ll E_{so}$ is respected. We also performed a numerical simulation of a 2D system on a rectangular lattice. We found that the pinning behavior of the lowest ABS is also reproduced in the system without the rotational invariance. Moreover, we perform numerical simulations of a two-dimensional heterostructure with a superconducting vortex, and we find that in such a setup with SOI one can also observe the formation of near zero-energy ABSs in the topologically trivial phase.
It will thus remain an experimental challenge to distinguish topological from non-topological bound states in standard transport measurements.

{\it Acknowledgements.} We acknowledge helpful discussions with Dmitry Miserev and Yanick Volpez. This work was supported by the Swiss National Science Foundation and NCCR QSIT. This project received funding from the European Union’s Horizon 2020 research and innovation program (ERC Starting Grant, grant agreement No 757725).


\begin{thebibliography}{99}
\bibitem{BesselJ}{The Bessel functions of the first kind are the solutions $f(x)$ of the differential equation $\left(x^2 d^2_{x} + x d{x} - \lambda^2+ k^2 x^2\right) f(x) $ $=$ $0$, where $f(x)$ $ =$ $ J_\lambda(k x)$, and are finite at $x = 0$.}
\bibitem{BesselK}{The modified Bessel functions of the second kind are the solutions $f(x)$ of the differential equation $\left(x^2 d^2_{x} +x d_{x} - \lambda^2- k^2 x^2\right) f(x)$ $=$ $0$, where $f(x)$ $=$ $ K_\lambda(kx)$, and are exponentially decaying at $x\rightarrow +\infty$ for $\Re{k}>0$.}
\end{thebibliography}
\end{document}